\documentclass[11pt,twoside]{JHEP3}
\usepackage{graphicx}
\title{Mass of a quantum 't~Hooft-Polyakov monopole}
\author{Arttu Rajantie\\
Theoretical Physics,
Blackett Laboratory, Imperial College, London SW7 2AZ, UK\\
E-mail: \email{a.rajantie@imperial.ac.uk}}
\abstract{The quantum mechanical mass of 't~Hooft-Polyakov monopoles in
the four-dimensional Georgi-Glashow is calculated non-perturbatively using lattice Monte Carlo
simulations. This is done by imposing twisted boundary conditions that ensure there is one unit of magnetic charge on the lattice, and measuring the free energy difference between this ensemble and the vacuum. In the weak-coupling limit, the results can be used to determine the quantum correction to the classical mass, once renormalisation of couplings is taken properly into account. 
The methods can also be used to study the masses at strong coupling, i.e., near the critical point, where there are hints of a possible electric-magnetic duality.
}

\begin{document}
\section{Introduction}
't~Hooft-Polyakov monopoles \cite{'tHooft:1974qc,Polyakov:1974ek} are topological solitons in the Georgi-Glashow model~\cite{Georgi:1972cj} and a wide range of other gauge field theories, 
including super Yang Mills theories and grand unified theories. They are non-linear objects in which energy is localised around a point in space and which therefore appear as point particles,
and they carry non-zero magnetic charge. It is possible that these monopoles actually exist in nature, but so far they
have not been discovered\footnote{One candidate event~\cite{Cabrera:1982gz} was seen on Valentine's Day 1982 but is unlikely to have been a real monopole.} despite extensive searches~\cite{Milton:2001qj}. However, 't~Hooft-Polyakov monopoles are very important theoretically, because they provide a new way of looking at non-Abelian gauge field theories, complementary to the
usual perturbative picture. In particular, this has shed more light on the puzzle of confinement~\cite{Mandelstam:1974pi,tHooftTalk}. 
So far, concrete results
have been limited to supersymmetric theories.

The main reason for the lack of progress in non-supersymmetric theories is the difficulty of treating the quantum corrections to the classical monopole solution. For instance, calculating the quantum correction to 
a soliton mass is a complicated task. Even in simple one-dimensional models, it can typically only
be calculated to one-loop order~\cite{Dashen:1974cj}, and for 't~Hooft-Polyakov monopoles the situation is even worse as only the leading logarithm is known~\cite{Kiselev:1988gf}. 
This difficulty is avoided in supersymmetric models, because the symmetry protects the mass from
quantum corrections.

In this paper, the quantum mechanical mass of a 't~Hooft-Polyakov monopole is calculated using lattice Monte Carlo simulations. The method was developed in Ref.~\cite{Davis:2000kv} and has been used earlier~\cite{Davis:2001mg} in a 2+1-dimensional
model in which the monopoles are instanton-like space-time events rather than particle excitations. The mass is defined using the free-energy difference between sectors with magnetic charges one and zero, and the corresponding ensembles are constructed using suitably twisted boundary conditions. This method has 
several advantages over the alternative approaches based on creation and annihilation operators~\cite{Frohlich:1998wq,Belavin:2002em,Khvedelidze:2005rv}
or fixed boundary conditions~\cite{Smit:1993gy,Cea:2000zr}. In particular, it gives a unique, unambiguous result, since it requires neither
gauge fixing, choice of a classical field configuration nor 
identification of individual monopoles in the field configurations.

Analogous twisted boundary conditions have been used before to compute
soliton masses in simpler models, such as 1+1-dimensional scalar field theory~\cite{Ciria:1993yx},
3+1-dimensional compact U(1) gauge theory~\cite{Vettorazzo:2003fg} and 2+1-dimensional Abelian Higgs model~\cite{Kajantie:1998zn}. In the latter case, the results provided evidence for an asymptotic duality near the critical point~\cite{Kajantie:2004vy}: The model becomes equivalent to a 
scalar field theory with a global O(2) symmetry, with vortices and scalar fields changing places.
It is interesting to speculate whether an electric-magnetic duality might appear in the same way in the Georgi-Glashow model.
These methods can, in principle, used to test that conjecture.

The outline of this paper is the following: The Georgi-Glashow model and the classical 't~Hooft-Polyakov solution are introduced in Section~\ref{sect:model}. In Section~\ref{sect:lattice}, the model is discretised on the lattice and the lattice magnetic field is defined. The twisted boundary conditions are discussed in Section~\ref{sect:twist}.
In Sections~\ref{sect:classmass} and \ref{sect:simu} the classical and quantum mechanical monopole masses are computed, and the results are discussed in Section~\ref{sect:discuss}. Finally, conclusions are presented in Section~\ref{sect:conclude}.

\section{Georgi-Glashow model}
\label{sect:model}
The 3+1-dimensional
Georgi-Glashow model~\cite{Georgi:1972cj} consists of an SU(2) gauge field $A_\mu$ and an Higgs field 
$\Phi$ in the adjoint representation, with the Lagrangian
\begin{equation}
{\cal L}=-\frac{1}{2}{\rm Tr}F_{\mu\nu}F^{\mu\nu}
+{\rm Tr}[D_\mu,\Phi][D^\mu,\Phi]-m^2{\rm Tr}\Phi^2-\lambda\left({\rm Tr}\Phi^2\right)^2,
\end{equation}
where the covariant derivative $D_\mu$ and the field strength tension are defined as $D_\mu=\partial_\mu+igA_\mu$ and $F_{\mu\nu}=[D_\mu,D_\nu]/ig$. $A_\mu$ and $\Phi$ are traceless, self-adjoint $2\times 2$ matrices, they can be
represented as linear combinations of Pauli $\sigma$ matrices,
\begin{equation}
\sigma_1=\left(\matrix{0 & 1 \cr 1 & 0}\right),\quad
\sigma_2=\left(\matrix{0 & -i \cr i & 0}\right),\quad
\sigma_3=\left(\matrix{1 & 0 \cr 0 & -1 }\right),
\end{equation}
as $A_\mu=A_\mu^a\sigma^a$, $\Phi=\Phi^a\sigma^a$.

On classical level, the model
has two dimensionless parameters, the coupling constants $g$ and $\lambda$, and
the scale is set by $m^2$. When $m^2$ is negative, the SU(2) symmetry is broken spontaneously to U(1) by a non-zero vacuum expectation value of the Higgs field ${\rm Tr}\Phi^2=v^2/2\equiv|m^2|/2\lambda$.
In the broken phase, the particle spectrum consists of a massless photon, electrically 
charged $W^{\pm}$ bosons with mass $m_W=gv$, a neutral Higgs scalar with mass $m_H=\sqrt{2\lambda}v$ and massive magnetic monopoles~\cite{'tHooft:1974qc,Polyakov:1974ek}.
The terms "electric" and "magnetic" refer to the effective U(1) field strength tensor defined 
as~\cite{'tHooft:1974qc}
\begin{equation}
\label{equ:fieldstrength}
{\cal F}_{\mu\nu}={\rm Tr}\hat\Phi F_{\mu\nu}-\frac{i}{2g}{\rm Tr}
\hat\Phi[D_\mu,\hat\Phi][D_\nu,\hat\Phi].
\end{equation}
In any smooth field configuration,
the corresponding magnetic field ${\cal B}_i=\epsilon_{ijk}{\cal F}_{jk}/2$ is sourceless (i.e., $\vec\nabla\cdot
\vec{\cal B}=0$) whenever $\Phi\ne 0$. This is easy to see in the unitary gauge,
in which $\Phi\propto\sigma_3$, because Eq.~(\ref{equ:fieldstrength}) reduces to 
${\cal F}_{\mu\nu}=\partial_\mu A^3_\nu-\partial_\nu A^3_\mu$ and therefore $\vec{\cal B}=\vec\nabla
\times\vec{A}^3$.
At zeros of $\Phi$, the divergence is $\pm 4\pi/g$ times a delta function, indicating a magnetic charge
of $q_M=4\pi/g$.

The classical 't~Hooft-Polyakov monopole solution~\cite{'tHooft:1974qc,Polyakov:1974ek} is of the form
\begin{eqnarray}
\Phi^a&=&\frac{r_a}{gr^2}H(gvr),
\nonumber\\
A_i&=&-\epsilon_{aij}\frac{r_j}{gr^2}\left[1-K(gvr)\right],
\end{eqnarray}
where $H(x)$ and $K(x)$ are functions that have to be determined numerically.
It is easy to check that this solution is a magnetic charge in the above sense. Because the energy
is localised around the origin, the solution describes a particle. Once the functions 
$H(x)$ and $K(x)$ have been found,
it is easy to integrate the energy functional to calculate the mass of the particle, as it is
simply given by the total energy of the configuration. The energy density
falls as $\rho\sim 1/2g^2r^4$, implying that the mass is finite but also that there is 
a long-range magnetic Coulomb force between monopoles, as expected.

The classical monopole mass $M_{\rm cl}$ can be written as
\begin{equation}
\label{equ:classmass}
M_{\rm cl}=\frac{4\pi m_W}{g^2}f(z),
\end{equation}
where $f(z)$ is a function of $z=m_H/m_W$ and is known to satisfy $f(0)=1$~\cite{Bogomolny:1975de,Prasad:1975kr}.
It has recently been calculated numerically to a high accuracy~\cite{Forgacs:2005vx}. 
Asymptotic expressions for small and large $z$ had
been found earlier~\cite{Kirkman:1981ck,Gardner:1982fk}, but the authors of Ref.~\cite{Forgacs:2005vx}
reported that they had found an error in the small-$z$ expansion. According to them,
the correct expansion is
\begin{equation}
f(z)=1+\frac{1}{2}z+\frac{1}{2}z^2\left(
\ln3\pi z-\frac{13}{12}-\frac{\pi^2}{36}
\right)+O(z^3).
\end{equation}
For large $z$, they found that
\begin{equation}
f(z)=1.7866584240(2)-2.228956(7)z^{-1}+7.14(1)z^{-2}+O(z^{-3}).
\end{equation}

In quantum theory, the mass of a soliton can be defined as the difference between the ground state energies in the sectors with one and zero charge. In principle, it is possible to calculate this
perturbatively to leading order~\cite{Dashen:1974cj}. First, one needs to find the classical solution
$\phi_0(x)$, and consider small fluctuations $\delta(t,x)$ around it,
\begin{equation}
\phi(t,x)=\phi_0(x)+\delta(t,x).
\end{equation}
When one drops higher-order terms in the Lagrangian, one is left with field $\delta$ in an harmonic $x$-dependent
potential $U(\delta)=\frac{1}{2}V''(\phi_0(x))\delta^2$.
One needs to find the energy levels $\omega_k$ of this field by solving the eigenvalue equation
\begin{equation}
\left[-\vec\nabla^2+V''(\phi_0(x))\right]\delta_k(x)=\omega_k^2\delta_k(x).
\end{equation}
The one-loop correction to the soliton mass is then given simply by the difference in the zero-point 
energies of one- and zero-soliton sectors,
\begin{equation}
M_{\rm 1-loop}=M_{\rm cl}+\frac{1}{2}\sum_k\left(\omega^1_k-\omega^0_k\right),
\end{equation}
where $\omega^1_k$ refers to the energies in the soliton background and $\omega^0_k$ in the trivial vacuum.

One has to be careful with degeneracies and 
ultraviolet divergences, but the calculation can be carried out exactly in, for 
instance, the 1+1-dimensional $\lambda\phi^4$ model. In the presence of a kink, the 
energy spectrum consists of two discrete levels $\omega_0^2=0$ and $\omega_1^2=(3/2)m^2$,
and a continuum $\omega_q^2=(q^2/2+2)m^2$. It is essential that one takes into account the same number of eigenvalues in the two sectors, and the best way to ensure is to do the calculation on a finite lattice and take the lattice spacing to zero and the lattice volume to infinity afterwards. This gives the result~\cite{Dashen:1974cj}
\begin{equation}
M_{\rm kink}=\frac{2\sqrt{2}}{3}\frac{m^3}{\lambda}\left[
1+\left(\frac{\sqrt{3}}{8}-\frac{9}{4\pi}\right)\frac{\lambda}{m^2}+O\left(\frac{\lambda^2}{m^4}\right)
\right].
\end{equation}

The one-loop calculation of the monopole mass would go along the same lines, but there are many extra complications, which make it technically more difficult. Instead of one field, one has to consider two coupled fields. The background solution is not known analytically except in the special case of $\lambda=0$, and even then, the eigenvalue equation cannot be solved analytically. It is difficult to regularise the theory without breaking either gauge or rotation invariance, and in any case, the theory has much stronger
ultraviolet divergences. Nevertheless, because the monopole mass is a physical quantity, it is finite
once the couplings have been renormalised. No separate renormalisation of the monopole mass is
needed, and that means that the scale dependence of the resulting one-loop expression for the monopole mass would automatically be such that it cancels the running of the couplings. 
 
So far, only the leading logarithmic quantum correction near the BPS limit has been calculated~\cite{Kiselev:1988gf},
\begin{equation}
\label{equ:leadinglog}
M=\frac{4\pi m_W}{g^2}\left[
1+\frac{g^2}{8\pi^2}\left(\ln\frac{m_H^2}{m_W^2}+O(1)
\right)+O(g^4)\right].
\end{equation}
An interesting aspect of this result is that the it is logarithmically divergent in the BPS limit.
This is related to the Coleman-Weinberg effect~\cite{Kiselev:1990fh}, which makes it impossible to actually
reach the BPS limit in the quantum theory.
Quantum corrections give rise to a logarithmic term $\phi^4\log\phi$, which means that if one tries
to decrease the scalar mass below a certain value, the vacuum becomes unstable. This leads to a 
constraint $m_H\gtrsim gm_W$. If one wants to be able to test Eq.~(\ref{equ:leadinglog}) numerically,
the logarithmic term has to be much larger than the constant term next to it, 
but that is only possible if $g$
is very small. This, however, means that the whole quantum correction will be small and therefore more
difficult to measure. Furthermore, having a large mass hierarchy such as $m_H\ll m_W$ means that 
one would need to use a very large lattice. For these reasons, a numerical test of Eq.~(\ref{equ:leadinglog}) is not attempted in this paper.

\section{Lattice discretisation}
\label{sect:lattice}
To study the model numerically, let us carry out a Wick rotation to Euclidean space and 
discretise the model in the standard way,
\begin{eqnarray}
\label{equ:latticeaction}
{\cal L}_E & = &
2\sum_{\mu}
\left[
{\rm Tr} \Phi(\vec{x})^2-
{\rm Tr} \Phi(\vec{x}) U_\mu(\vec{x}) \Phi(\vec{x}+\hat{\mu})
U_\mu^\dagger(\vec{x})\right]
\nonumber
\\
& &+\frac{2}{g^2}\sum_{\mu<\nu}\left[2-
{\rm Tr} U_{\mu\nu}(\vec{x})
\right]
+m^2{\rm Tr}\ \Phi^2+\lambda({\rm Tr}\ \Phi^2)^2.
\end{eqnarray}
The scalar field $\Phi$ is defined on lattice sites and the gauge field is represented by
SU(2)-valued link variables $U_\mu$, which correspond roughly to $\exp(igA_\mu)$. The plaquette $U_{\mu\nu}$ is defined as $U_{\mu\nu}(\vec{x})=U_\mu(\vec{x})U_\nu(\vec{x}+\hat\mu)
U^\dagger_\mu(\vec{x}+\hat\nu)U^\dagger_\nu(\vec{x})$.

It will be crucial that the magnetic field can be defined on the lattice and that magnetic monopoles are therefore absolutely stable objects~\cite{Davis:2000kv}. This is highly non-trivial, 
because many other topological objects such as Yang-Mills instantons are not well defined on 
the lattice~\cite{Luscher:1981zq}. 
To define the discretised version of the field strength tensor ${\cal F}_{\mu\nu}$,
note that the set of configurations with $\Phi=0$ at any lattice site is of measure zero and 
therefore these configurations do not contribute to any physical observables. 
One can therefore define a unit vector valued field $\hat\Phi=\Phi/\sqrt{\Phi^2}$. This expression makes sense because $\Phi^2$ is always proportional to the $2\times 2$ identity matrix ${\mathbf 1}$. 

Because $\hat\Phi^2={\mathbf 1}$, one 
can define a projection operator $\Pi_+=({\mathbf 1}+\hat\Phi)/2$.
Let us use it to define the projected link variable
\begin{equation}
u_\mu(x)=\Pi_+(x) U_\mu(x)
\Pi_+(x+\hat\mu)
,
\end{equation}
which is essentially the compact Abelian gauge field that corresponds to the unbroken U(1) subgroup.
The corresponding Abelian field strength tensor is
\begin{equation}
\alpha_{\mu\nu}=\frac{2}{g}\arg{\rm Tr}\,
u_\mu(x)u_\nu(x+\hat\mu)
u^\dagger_\mu(x+\hat\nu)u^\dagger_\nu(x),
\end{equation}
and the lattice version of the magnetic field
\begin{equation}
\label{equ:latB}
\hat B_i=\frac{1}{2}\epsilon_{ijk}\alpha_{jk}.
\end{equation}

The lattice magnetic field $\hat{B}_i$ is a well-defined, gauge-invariant quantity. The magnetic charge density is given by its divergence,
\begin{equation}
\rho_M(x)=\sum_{i=1}^3\left[\hat{B}_i(x+i)-\hat{B}_i(x)\right]\in \frac{4\pi}{g}{\mathbb Z},
\end{equation}
and note that it is quantised. Being a divergence of a vector field, the magnetic charge is automatically
a conserved quantity. 

\section{Twisted boundary conditions}
\label{sect:twist}
Because the magnetic charge $Q_M=\int d^3x \rho_M$ defined in the previous section is a well-defined, gauge-invariant,
quantised and conserved quantity even on the lattice, it is a well-defined question to
ask what the lowest energy eigenvalue with $Q_M=4\pi/g$ is. Furthermore, since the total magnetic
charge inside a volume is given by a surface integral over the boundary, one can fix the 
total charge in a simulation by choosing appropriate boundary conditions.

In practice, one can therefore 
define separate partition functions $Z_{N}$ for each magnetic charge sector,
\begin{equation}
Z_{N}=\int_{N} DU_\mu D\Phi \exp(-S[U_\mu,\Phi]),
\end{equation}
where the boundary conditions for each sector are such that they fix the magnetic charge to $Q_M=4\pi N/g$, i.e., the net number of monopoles is $N$.
Since monopoles of the same charge are not expected to form bound states, and since their
interaction potential decreases with distance as $1/r$, they will be non-interacting
provided that the lattice is large enough. Denoting the length of the time direction by $T$, the
partition function is therefore
\begin{equation}
Z_N=\exp(-|N|MT)Z_0,
\end{equation}
where $M$ is the quantum mechanical mass of the monopole, and $Z_0$ is the partition function with $N=0$.
In particular, one can express the mass as
\begin{equation}
M=-\frac{1}{T}\ln\frac{Z_1}{Z_0}.
\end{equation}

It was shown in Ref.~\cite{Davis:2000kv} that this can be achieved by using suitably "twisted" boundary
conditions. It is clear that periodic boundary conditions will not be useful, because they will fix
the total charge to zero. On the other hand, they have the attractive feature that they preserve translation
invariance and therefore, as all lattice points are equivalent, there will be no physical boundary.
However, this does not require perfect periodicity: Periodicity up to the symmetries of the Lagrangian
is enough. An obvious alternative is to use C-periodic boundary conditions~\cite{Kronfeld:1990qu},
\begin{eqnarray}
\label{equ:Cper}
U_\mu(x+N\hat\jmath)&=&U_\mu^*(x)=\sigma_2U_\mu(x)\sigma_2,\nonumber\\
\Phi(x+N\hat\jmath)&=&\Phi^*(x)=-\sigma_2\Phi(x)\sigma_2.
\end{eqnarray}
They flip the sign of the magnetic field as one goes through the boundary and therefore 
allow non-zero magnetic charges. 

In fact, it turns out that C-periodic boundary conditions
allow the magnetic charge to have any even value~\cite{Davis:2000kv}. This means that the one-monopole sector is 
still excluded, and also that in practice the magnetic charge will always be zero, because as long as
$M$ is non-zero and $T$ is chosen to be large enough (i.e., $T\gg 1/M$),
\begin{equation}
Z_{\rm C}=\sum_{k=-\infty}^\infty Z_{2k}=Z_0\left(1+O(e^{-2MT})\right).
\end{equation}

When the boundary conditions in Eq.~(\ref{equ:Cper}) are written in the matrix form, it becomes obvious 
that one could have used $\sigma_1$ or $\sigma_3$ instead of $\sigma_2$. They are all related to each other
by gauge transformations 
and therefore describe identical physical situations. However, this observation
allows one to define twisted boundary conditions
\begin{eqnarray}
\label{equ:twisted}
U_\mu(x+N\hat\jmath)&=&\sigma_jU_\mu(x)\sigma_j,\nonumber\\
\Phi(x+N\hat\jmath)&=&-\sigma_j\Phi(x)\sigma_j,
\end{eqnarray}
which are locally equivalent to Eq.~(\ref{equ:Cper}), but not globally. It is possible to carry out
a gauge transform to turn the boundary conditions to Eq.~(\ref{equ:Cper}) in any single direction,
but it is not possible  to do it to all three directions simultaneously. Considering the total
charge of the lattice, one finds that these twisted boundary conditions only allow odd 
values~\cite{Davis:2000kv}, 
and therefore
\begin{equation}
Z_{\rm tw}=\sum_{k=-\infty}^\infty Z_{2k+1}=Z_1\left(2+O(e^{-2MT})\right).
\end{equation}
Thus, the ratio of the twisted and C-periodic boundary conditions can be used to calculate the monopole mass,
\begin{equation}
\label{equ:massdef}
-\frac{1}{T}\ln\frac{Z_{\rm tw}}{Z_{\rm C}}=M-\frac{\ln 2}{T}+O(e^{-2MT})\to M
\quad\mbox{as $T\rightarrow\infty$}.
\end{equation}

As such, this expression is of little use, because
it is not possible measure partition functions directly in Monte Carlo simulations. One cannot
write the ratio of partition functions in Eq.~(\ref{equ:massdef}) as an expectation value either,
because $Z_{\rm tw}$ and $Z_{\rm C}$ have different boundary conditions. One
possible way to avoid this problem is to change the integration variables in $Z_{\rm tw}$
in such a way that the new variables satisfy C-periodic boundary conditions. This changes the integrand,
or equivalently the action $S\rightarrow S+\Delta S$. This way, one can express Eq.~(\ref{equ:massdef})
in terms of an expectation value
$Z_{\rm tw}/Z_{\rm C}=\langle\exp(-\Delta S)\rangle_{\rm C}$, where the subscript ${\rm C}$ indicates that 
the expectation value is calculated with C-periodic boundary conditions.
In principle, this is measurable in the simulations.
The shift $\Delta S$ consists of line integrals of the magnetic field around the lattice~\cite{Davis:2000kv}.
In practice, this approach does not work, because $\exp(-\Delta S)$ has very little overlap with the vacuum and one would need extremely high statistics to obtain any meaningful results. 

Let us, however, adopt a different strategy. Going back to Eq.~(\ref{equ:massdef}), we can differentiate the mass with respect to some parameter $x$,
\begin{equation}
\frac{\partial M}{\partial x}=\frac{1}{T}\left(
\left\langle\frac{\partial S}{\partial x}\right\rangle_{\rm tw}
-
\left\langle\frac{\partial S}{\partial x}\right\rangle_{\rm C}
\right),
\end{equation}
where the subscripts "tw" and "C" refer to expectation values calculated with twisted and C-periodic 
boundary conditions, respectively.
If one start at a point where one knows the monopole mass, one can integrate this to obtain the mass at
any other parameter values. Possible choices for the start point of the integration are the classical 
limit, where $M$ can be calculated directly, or any point in the symmetric phase where the monopole mass vanishes.

Let us choose the latter option and use $x=m^2$. Thus we can write
\begin{equation}
\label{equ:massderiv}
\frac{\partial M}{\partial m^2}=L^3\left(
\left\langle{\rm Tr}\Phi^2\right\rangle_{\rm tw}
-
\left\langle{\rm Tr}\Phi^2\right\rangle_{\rm C}
\right).
\end{equation}
If one chooses a large enough initial value for $m^2$, it is guaranteed to be in the symmetric phase.
In fact, since one can only carry out a finite number of measurements, it is better to use finite differences instead of
derivatives
\begin{equation}
\label{equ:finitediff}
M(m^2_2)-M(m^2_1)=-\frac{1}{T}\ln
\frac{\left\langle e^{-(m^2_2-m^2_1)\sum_x {\rm Tr}\Phi^2}\right\rangle_{m^2_1,\rm tw}}{
\left\langle e^{-(m^2_2-m^2_1)\sum_x {\rm Tr}\Phi^2}\right\rangle_{m^2_1,\rm C}},
\end{equation}
where the subscript $m^2_1$ indicates that the expectation value is calculated at $m^2=m^2_1$.
The spacing between different values of $m^2$ has to be fine enough so that the expectation values
can be calculated reliably.

\section{Classical Mass}
\label{sect:classmass}
It will be interesting to compare the measured quantum masses with classical results 
to determine the quantum correction. However, the quantum mass will be computed on a finite lattice,
and therefore it does not make sense to compare it with the infinite-volume continuum expression 
(\ref{equ:classmass}). Instead, one needs to know the classical mass on the same finite lattice. That is
straightforward to compute by minimising the lattice action $S_{\rm tw}$ (\ref{equ:latticeaction}) with twisted boundary conditions. The C-periodic boundary conditions
are compatible with the classical vacuum solution, and therefore the minimum action in that sector is
simply $S_{\rm C}^{\rm min}=-(m^4/4\lambda)TL^3$. The classical mass is therefore given by
\begin{equation}
\label{equ:Sdiff}
M_{\rm cl}=\frac{S_{\rm tw}^{\rm min}-S_{\rm C}^{\rm min}}{T}=
\frac{S_{\rm tw}^{\rm min}}{T}+\frac{m^4}{4\lambda}L^3.
\end{equation}
In fact, since the classical monopole configuration is time-independent, it is enough to have
only one time step in the time direction, $T=1$.

\FIGURE{
\includegraphics[width=10cm]{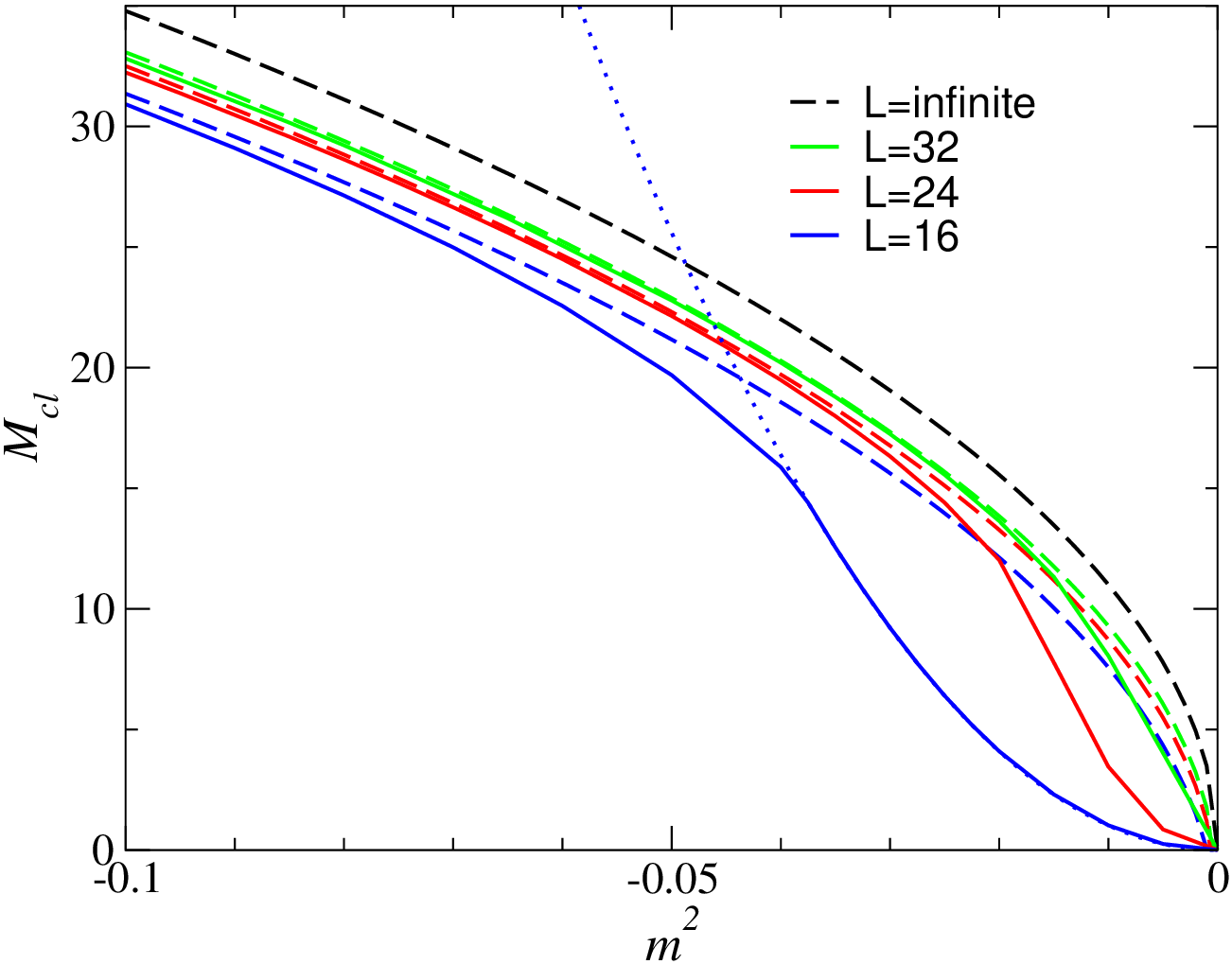}
\caption{\label{fig:classmass}Classical monopole mass.}
}

The classical monopole mass was measured on three different lattices, $16^3$, $24^3$ and $32^3$ using 
couplings $\lambda=0.1$ and $g=1/\sqrt{5}$, which correspond to $z=1$. The results are shown in Fig.~\ref{fig:classmass}. The coloured solid lines correspond to different lattice sizes, the smallest
being at the bottom. The top dashed line (black) is the infinite-volume mass given by Eq.~(\ref{equ:classmass}) 
with $f(1)\approx 1.238$ as computed in Ref.~\cite{Forgacs:2005vx}.
These results show a significant finite-size effect and demonstrate why it is 
necessary to compare the quantum result with the classical mass on the same lattice.

Deep in the broken phase, where $m^2\ll 0$, the finite-size effect should be due to the magnetic Coulomb
interaction between monopoles. Because our boundary conditions have the physical effect of
charge conjugation, we effectively have monopoles and antimonopoles alternating in a cubic
array, with distance $L$ between them. The energy of such a configuration is
\begin{equation}
E(L)=M-\frac{2\pi}{g^2L}\sum_{\vec{n}\ne 0}\frac{1}{|\vec{n}|}\approx 
M-\frac{10.98}{g^2L}.
\label{equ:Coulombeffect}
\end{equation}
The lower dashed lines (coloured) 
show the predicted finite-size effects for the relevant lattice volumes,
and one can see that the agreement is good deep in the broken phase. In fact, the lattice values are
slightly below the continuum results based on Ref.~\cite{Forgacs:2005vx}. This is most likely due to
discretisation effects.

Although the Coulomb interaction describes the finite-size effects very well deep in the broken phase, it 
fails badly as $m^2\rightarrow 0$. What happens there is that the size of the monopole, which is
proportional to $1/\sqrt{|m^2|}$, grows and eventually becomes comparable with the size of the lattice.
At some point it becomes energetically favourable for the whole system to remain in the symmetric phase. 
Because the field $\Phi$ is zero, the twisted action $S_{\rm tw}^{\rm min}$ in Eq.~(\ref{equ:Sdiff})
vanishes, and the result is
\begin{equation}
E_{\rm symm}(L)=V(0)L^3=\frac{m^4L^3}{4\lambda}.
\end{equation}
This is shown as a dotted line for $L=16$ in Fig.~\ref{fig:classmass}, and agrees well with the result near $m^2=0$. At intermediate values of $m^2$, the minimum energy configuration corresponds to
a deformed monopole, and therefore the actual result interpolates smoothly between the two behaviours.
Nevertheless, we have identified the main sources of finite-size 
effects in the classical calculation, and we
are therefore in a position to compare quantum and classical calculations.

\section{Simulations}
\label{sect:simu}
In the quantum simulations, the ensembles of
configurations with twisted and C-periodic boundary conditions were generated using the Metropolis algorithm. Three different lattice sizes were used:
$16^4$, $24^3\times 16$ and $32^3\times 16$. 
The system was first equilibrated by carrying out 20000--60000 update sweeps depending on the lattice size and the value of $m^2$, and after that, measurements were carried out every 100 updates. The number of measurements for each value of $m^2$ was between 100 and 1700.

\vspace*{1cm}

\FIGURE{
\includegraphics[width=10cm]{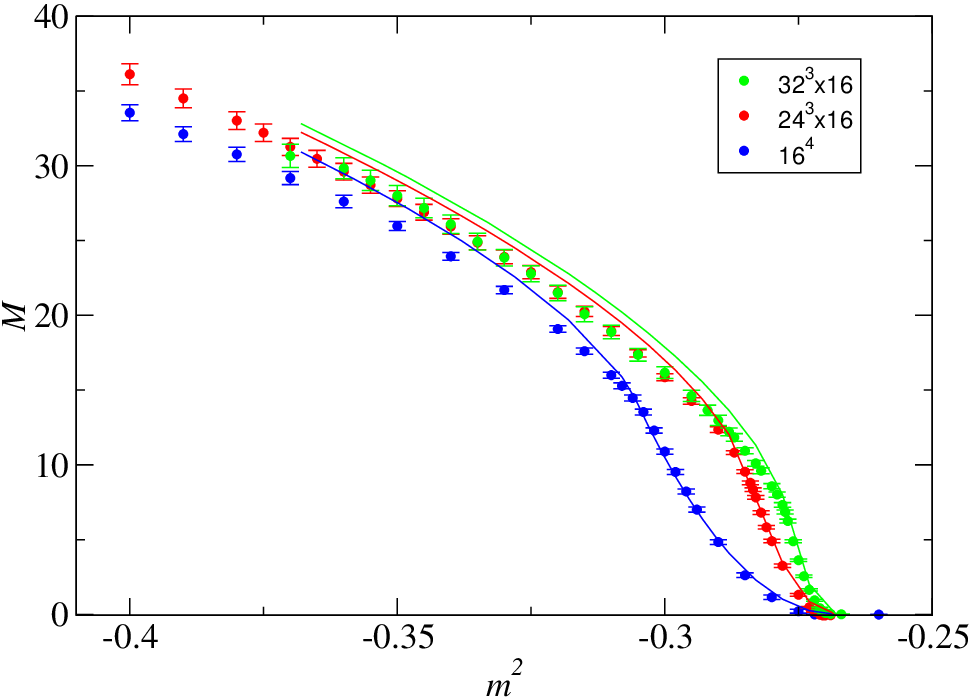}
\caption{\label{fig:quantummass}Quantum monopole mass (points) compared with the classical mass (lines).}
}

The expectation values needed for the change of $M$ from $m^2_1$ to $m^2_2$ can be calculated in two 
different ways,
\begin{equation}
\left\langle e^{-(m^2_2-m^2_1)\sum_x {\rm Tr}\Phi^2}\right\rangle_{m^2_1}
=
\left\langle e^{-(m^2_1-m^2_2)\sum_x {\rm Tr}\Phi^2}\right\rangle_{m^2_2}.
\end{equation}
This was be used to check that the system was properly in equilibrium, the statistics were
sufficient and the spacing between different values of $m^2$ was small enough. Defining
\begin{equation}
f_1=-\frac{1}{T}\ln\left\langle e^{-(m^2_2-m^2_1)\sum_x {\rm Tr}\Phi^2}\right\rangle_1\quad \mbox{and}\qquad
f_2=-\frac{1}{T}\ln\left\langle e^{-(m^1_2-m^2_1)\sum_x {\rm Tr}\Phi^2}\right\rangle_2,
\end{equation}
the change in the monopole mass (\ref{equ:finitediff}) can be written as
\begin{equation}
M(m_2^2)-M(m_1^2)=\frac{1}{2}\left(f_{1,{\rm tw}}+f_{2,{\rm tw}}-f_{1,{\rm C}}-f_{2,{\rm C}}\right).
\end{equation}
The statistical error $\Delta f$ in each $f_{i,X}$ was estimated using the bootstrap method and it was 
made sure that $f_{1,X}$ and $f_{2,X}$ agreed within the errors.
The error in the mass difference was estimated to be
\begin{eqnarray}
\Delta\left[M(m_2^2)-M(m_1^2)\right]^2
&=&
\frac{1}{4}\left[
\Delta f_{1,{\rm tw}}^2+\Delta f_{2,{\rm tw}}^2+\left(f_{1,{\rm tw}}-f_{2,{\rm tw}}\right)^2
\right.\nonumber\\&&\left.
+
\Delta f_{1,{\rm C}}^2+\Delta f_{2,{\rm C}}^2+\left(f_{1,{\rm C}}-f_{2,{\rm C}}\right)^2
\right]
\end{eqnarray}
The differences were then summed up, starting from $m_0^2$, the highest value of $m^2$, where $M$ was assumed to be zero, 
\begin{equation}
M(m_N^2)=\sum_{n=0}^{N-1} \left(M(m_{n+1}^2)-M(m_n^2)\right).
\end{equation}
The total error was calculated by assuming that the errors in the individual mass differences were independent,
\begin{equation}
\Delta[M(m_N^2)]^2=\sum_{n=0}^{N-1} \Delta\left[M(m_{n+1}^2)-M(m_n^2)\right]^2
\end{equation}
The results are shown in Fig.~\ref{fig:quantummass}. 
Note that the errors are highly correlated.
In Fig.~\ref{fig:massderiv}, we show the derivative of the mass calculated from Eq.~(\ref{equ:massderiv}).

\vspace*{1cm}

\FIGURE{
\includegraphics[width=10cm]{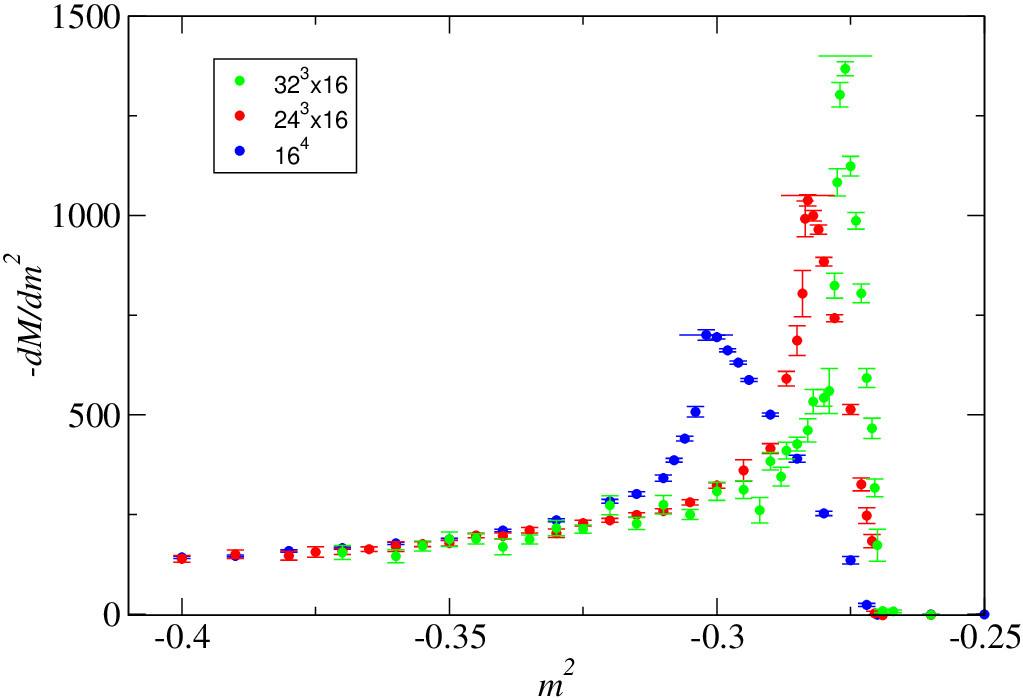}
\caption{\label{fig:massderiv}Derivative of the monopole mass.}}

\section{Discussion}
\label{sect:discuss}
The mass derivative in Fig.~\ref{fig:massderiv} has a sharp peak, above which it drops rapidly to zero. 
This is compatible with the classical result $\partial M/\partial m^2\propto 1/\sqrt{-m^2}$ for negative $m^2$ and zero for positive values.
As the horizontal lines show, the peak height is proportional to the linear size of the lattice,
which is what one would expect to happen in a second-order phase transition.
A fit to the peak position gives an infinite-volume value $m_c^2\approx 0.268$ for the critical point.

The curves in Fig.~\ref{fig:quantummass} show the classical results shifted by this amount. The quantum measurements agree fairly well with them near the critical point, but start to deviate deeper in the broken phase. The qualitative behaviour, as well as the finite-size effects, are nevertheless similar.

To really carry out a quantitative comparison of the quantum and classical results and to extract the quantum correction to the mass, one has to consider the renormalisation of the parameters.
The values of $m^2$, $\lambda$ and $g$ that were used in the simulations correspond to bare couplings, 
but one should compare
the measurements with the classical mass calculated using the corresponding renormalised couplings $m_R^2$, $\lambda_R$ and $g_R$. The values of the renormalised couplings depend on the renormalisation scheme
and scale, and therefore there is no unique way to compare the results.
Furthermore, if one calculates the renormalisation counterterms to a certain order in
perturbation theory, the value of quantum correction obtained by subtracting the classical value 
from the quantum result is only valid to the same order, even though the quantum mass itself
has been calculated fully non-perturbatively.

\vspace*{1cm}

\FIGURE{
\includegraphics[width=10cm]{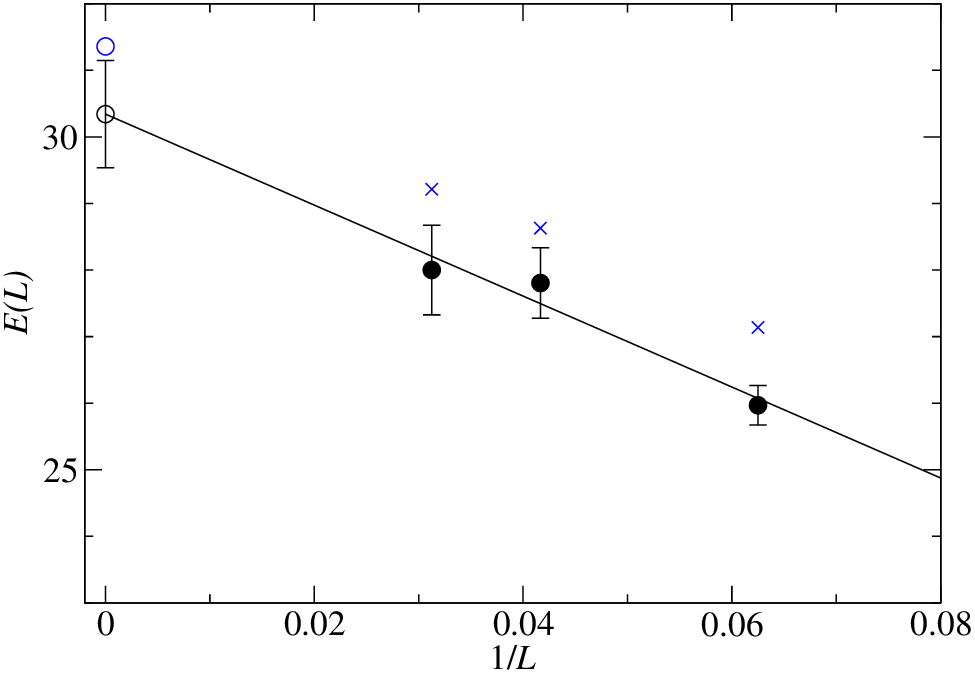}
\caption{\label{fig:gRdet}Measurements of the monopole mass (black circles) at $m^2=-0.35$ with different lattice sizes $L$. The solid line is a fit of the form (\ref{equ:Coulombeffect}) and gives $g_R=0.40(6)$. The blue crosses show the corresponding classical masses.}
}

It would, therefore, be best to use a physically meaningful renormalisation
scheme and compute the renormalised couplings non-perturbatively. This can be done by choosing
three observable quantities $X$, $Y$ and $Z$, one for each coupling, and measuring their values 
$\langle X\rangle$, $\langle Y\rangle$ and $\langle Z\rangle$
in Monte Carlo simulations. One would then calculate the same quantities in the classical theory,
and fix the values of the renormalised couplings by requiring that the classical values agree with the quantum measurements,
\begin{equation}
X_{\rm cl}(m_R^2,\lambda_R,g_R)=\langle X\rangle\quad\mbox{etc.}
\end{equation}
It would be natural to choose the masses of the perturbative excitations $m_H$ and $m_W$ as two of these observables, although measuring $m_W$ is non-trivial because of its electric charge. One can choose the monopole charge as the third observable, because its value can be determined relatively straightforwardly from the finite-size effects of the monopole mass. In Fig.~\ref{fig:gRdet} we show the measured finite-size effects at $m^2=-0.35$. A fit to Eq.~(\ref{equ:Coulombeffect}) with $g$ and $M$ as free parameters gives $g_R=0.40(6)$. It agrees with the bare value $g=1/\sqrt{5}\approx 0.447$ within errors, so
one would need better statistics to be actually able to measure the renormalisation counterterm.

The change of the monopole mass as a function of $m^2$ is also directly related to the renormalisation of the theory. In a classical continuum theory, $m^2$ only fixes the scale, and dimensionless observables do not depend on it. In the quantum theory, this scale invariance is broken, and taking $m^2$ towards the critical point corresponds to renormalisation group flow towards infrared. 
Roughly speaking one can identify $m_H$ with the renormalisation scale $\mu$.
In principle, one should therefore be able to use the non-perturbative renormalisation scheme discussed above to follow the running of the couplings even in the non-perturbative regime near the critical point.

One can speculate on what may happen based on the perturbative running of the couplings. 
The one-loop renormalisation group equations are
\begin{equation}
\frac{d\lambda_R}{d\log\mu}=\frac{11\lambda_R^2-12g_R^2\lambda_R+6g_R^4}{8\pi^2},
\qquad
\frac{dg_R^2}{d\log\mu}=-\frac{7g_R^4}{8\pi^2}.
\end{equation}
Moving towards infrared, $\lambda_R$ decreases and $g_R^2$ increases. 
In fact, $\lambda_R$ becomes
negative at a non-zero $\mu$, i.e., before one reaches the critical point. This is a sign of the 
Coleman-Weinberg effect and means that there is a first-order phase transition. However, if $g_R$
has become large enough before this happens, the one-loop approximation is not valid any more,
and it is possible that the critical point can be reached. This would mean that the line of
first-order Coleman-Weinberg phase transitions ends at a tricritical point. Beyond that there is a 
second-order phase transition, around which the theory is strongly coupled.

This is exactly what happens in the Abelian Higgs model in 2+1 dimensions. There are strong arguments
that in that case, the second-order phase transition has a dual description in terms of a global O(2) model~\cite{KleinertBook}. In the duality map, vortices and the fundamental scalar fields of the models change places. This was recently tested by measuring the critical behaviour of the vortex mass using a technique that was very similar to what has been discussed in this paper~\cite{Kajantie:2004vy}. The critical exponents that characterise that behaviour of vortices near the transition point were found to agree with the known critical exponents of the O(2) model, which provides strong numerical evidence for the duality.

If the Georgi-Glashow model has a second-order phase transition, it may have an analogous dual description. The one-loop renormalisation group equations suggest that as the critical point is approached, $g_R$ diverges and $\lambda_R$ goes to zero. The masses of the $H$ and $W^\pm$ bosons and the monopoles should behave as
\begin{equation}
m_H\propto\lambda_R^{1/2},\quad m_W\propto g_R, \quad M\propto g_R^{-1}, 
\end{equation}
implying that near the critical point, the $W^\pm$ bosons become much heavier than the other degrees of freedom and decouple. The Higgs scalar is neutral, and therefore it decouples as well. Thus one is left with massive magnetic monopoles coupled to a massless photon field. Because of the symmetry between electric and magnetic fields in electrodynamics, this system is indistinguishable from one with an electrically charged scalar field, i.e., the Abelian Higgs model.

It is therefore possible that near the critical point, the broken phase of the Georgi-Glashow model is dual to the symmetric phase of the Abelian Higgs model. If the duality extends through the phase transition, the confining symmetric phase of the Georgi-Glashow model is dual to the broken superconducting phase of the Abelian Higgs model, with Abrikosov flux tubes playing the role of the confining strings. This is a concrete example of the 't~Hooft-Mandelstam picture of confinement as a dual phenomenon to superconductivity~\cite{tHooftTalk,Mandelstam:1974pi}.

Earlier studies of the Georgi-Glashow model shed some light on this possible duality~\cite{Greensite:2004ke}. 
Best known is the limit $\lambda\rightarrow\infty$ taken with constant $v^2=|m^2|/\lambda$.
It corresponds to fixing the norm of the Higgs field $\Phi$, and is traditionally parameterised by couplings $\kappa=(m^2+8)/\lambda$ and $\beta=4/g^2$. The limits $\kappa\rightarrow\infty$ and $\beta\rightarrow\infty$ of that theory are, respectively, the compact U(1) gauge theory and the global O(3) spin model. The former is believed to have a weakly first-order phase transition~\cite{Vettorazzo:2003fg}, and the latter a second-order one. These two transitions are connected by a phase transition line that separates the Higgs and confining phases. There is evidence for a tricritical point at finite $(\kappa,\beta)$ at which the transition changes from first to second order~\cite{Baier:1988sc}, but it is not known if the tricritical line extends to $\lambda=0$.

Interestingly, the $\kappa=\infty$ limit, i.e., the compact U(1) theory, is exactly dual to the so-called frozen superconductor~\cite{Peskin:1977kp}. This is an Abelian integer-valued gauge theory, which can be obtained as the $\lambda\rightarrow\infty$, $\kappa\rightarrow\infty$ limit of the Abelian Higgs model. In other words, the hypothetical duality discussed above is real and exact in this particular limit of the theory. The interesting question is whether it exists as an asymptotic duality even away from the $\kappa=\infty$ limit. This is by no means clear because the compact U(1) theory does not have a second-order transition.

In principle, the methods discussed and used in this paper can be used to test the duality hypothesis. If one finds a second-order phase transition, one can measure how the masses and the gauge coupling $g_R$ change as the transition is approached and determine whether the $W^\pm$ particles decouple. One can then construct observables along the lines of Ref.~\cite{Kajantie:2004vy} to compare the critical behaviours of the monopoles in the Georgi-Glashow model and scalar particles in the Abelian Higgs model. This will, however, be a major computation, and is beyond the scope of this paper.

\section{Conclusions}
\label{sect:conclude}
We have seen how the quantum mechanical mass of a 't~Hooft-Polyakov monopole can be calculated non-perturbatively using twisted boundary conditions. The method has clear advantages over alternative approaches based on creation and annihilation operators and fixed boundary conditions.
While similar calculations have been carried out before in simpler models~\cite{Ciria:1993yx,Vettorazzo:2003fg,Kajantie:1998zn}, this appears to
be the first time it has been used for 't~Hooft-Polyakov monopoles in 3+1 dimensions.

The results demonstrate that one can obtain relatively accurate results for the monopole mass. It would be interesting to compare the results with the corresponding classical mass to determine the quantum correction. As we have seen, the finite-size effects due to the magnetic Coulomb interactions are significant, and therefore one has to compute the classical mass on the same lattice to have a meaningful comparison. Furthermore, the classical mass has to be calculated using the renormalised
rather than bare couplings, and this introduces a dependence on the renormalisation scheme and scale. 

The simulations in this paper were done at weak coupling, i.e., deep in the broken phase.
This is a useful limit for testing the method and also for identifying the quantum correction. However, the strong-coupling limit, which corresponds to the neighbourhood of the transition point, is arguably more interesting. In perturbation theory, the transition is of first order, and therefore one cannot reach the critical point, but it is possible that this changes if $\lambda$ is high enough. The methods discussed in this paper could then be used to study the critical behaviour.

A particularly interesting possibility is that an asymptotic electric-magnetic duality appears near the critical point. The theory would then become equivalent to the Abelian Higgs model, with monopoles playing the role of the charged scalars. This would be a concrete example of the picture of confinement as a dual phenomenon to superconductivity.

\acknowledgments
The author would like to thank Philippe de Forcrand, Tanmay Vachaspati 
and Falk Bruckmann 
for useful discussions. The research was conducted in cooperation with SGI/Intel utilising the Altix 3700 supercomputer and
was supported in part by Churchill College, Cambridge, and the National Science Foundation under Grant No. PHY99-07949.

\bibliography{monomass}

\begin{thebibliography}{100}
%\cite{'tHooft:1974qc}
\bibitem{'tHooft:1974qc}
  G.~'t Hooft,
  %``Magnetic Monopoles In Unified Gauge Theories,''
  Nucl.\ Phys.\ B {\bf 79} (1974) 276.
  %%CITATION = NUPHA,B79,276;%%
  
%\cite{Polyakov:1974ek}
\bibitem{Polyakov:1974ek}
  A.~M.~Polyakov,
  %``Particle Spectrum In Quantum Field Theory,''
  JETP Lett.\  {\bf 20} (1974) 194
  [Pisma Zh.\ Eksp.\ Teor.\ Fiz.\  {\bf 20} (1974) 430].
  %%CITATION = JTPLA,20,194;%%

%\cite{Georgi:1972cj}
\bibitem{Georgi:1972cj}
  H.~Georgi and S.~L.~Glashow,
  %``Unified Weak And Electromagnetic Interactions Without Neutral Currents,''
  Phys.\ Rev.\ Lett.\  {\bf 28} (1972) 1494.
  %%CITATION = PRLTA,28,1494;%%

%\cite{Cabrera:1982gz}
\bibitem{Cabrera:1982gz}
  B.~Cabrera,
  %``First Results From A Superconductive Detector For Moving Magnetic
  %Monopoles,''
  Phys.\ Rev.\ Lett.\  {\bf 48} (1982) 1378.
  %%CITATION = PRLTA,48,1378;%%
  
%\cite{Milton:2001qj}
\bibitem{Milton:2001qj}
  K.~A.~Milton, G.~R.~Kalbfleisch, W.~Luo and L.~Gamberg,
  %``Theoretical and experimental status of magnetic monopoles,''
  Int.\ J.\ Mod.\ Phys.\ A {\bf 17} (2002) 732
  [arXiv:hep-ph/0111062].
  %%CITATION = HEP-PH 0111062;%%

\bibitem{tHooftTalk}
G.~'t~Hooft, in {\it High Energy Physics}, EPS International Conference, Palermo, 1975, edited by A. Zichichi (Compositori, Bologna, 1976).
    
%\cite{Mandelstam:1974pi}
\bibitem{Mandelstam:1974pi}
  S.~Mandelstam,
  %``Vortices And Quark Confinement In Nonabelian Gauge Theories,''
  Phys.\ Rept.\  {\bf 23}, 245 (1976).
  %%CITATION = PRPLC,23,245;%%
  
%\cite{Dashen:1974cj}
\bibitem{Dashen:1974cj}
  R.~F.~Dashen, B.~Hasslacher and A.~Neveu,
  %``Nonperturbative Methods And Extended Hadron Models In Field Theory. 2.
  %Two-Dimensional Models And Extended Hadrons,''
  Phys.\ Rev.\ D {\bf 10} (1974) 4130.
  %%CITATION = PHRVA,D10,4130;%%
  
%\cite{Kiselev:1988gf}
\bibitem{Kiselev:1988gf}
  V.~G.~Kiselev and K.~G.~Selivanov,
  %``Quantum Correction To Monopole Mass,''
  Phys.\ Lett.\ B {\bf 213}, 165 (1988).
  %%CITATION = PHLTA,B213,165;%%  

%\cite{Davis:2000kv}
\bibitem{Davis:2000kv}
  A.~C.~Davis, T.~W.~B.~Kibble, A.~Rajantie and H.~Shanahan,
  %``Topological defects in lattice gauge theories,''
  JHEP {\bf 0011}, 010 (2000)
  [arXiv:hep-lat/0009037].
  %%CITATION = HEP-LAT 0009037;%%  
  
%\cite{Davis:2001mg}
\bibitem{Davis:2001mg}
  A.~C.~Davis, A.~Hart, T.~W.~B.~Kibble and A.~Rajantie,
  %``The monopole mass in the three-dimensional Georgi-Glashow model,''
  Phys.\ Rev.\ D {\bf 65}, 125008 (2002)
  [arXiv:hep-lat/0110154].
  %%CITATION = HEP-LAT 0110154;%%
  
%\cite{Frohlich:1998wq}
\bibitem{Frohlich:1998wq}
  J.~Frohlich and P.~A.~Marchetti,
  %``Gauge-invariant charged, monopole and dyon fields in gauge theories,''
  Nucl.\ Phys.\ B {\bf 551}, 770 (1999)
  [arXiv:hep-th/9812004].
  %%CITATION = HEP-TH 9812004;%%
  
%\cite{Belavin:2002em}
\bibitem{Belavin:2002em}
  V.~A.~Belavin, M.~N.~Chernodub and M.~I.~Polikarpov,
  %``Monopole creation operators as confinement-deconfinement order
  %parameters,''
  Phys.\ Lett.\ B {\bf 554}, 146 (2003)
  [arXiv:hep-lat/0212004].
  %%CITATION = HEP-LAT 0212004;%%

%\cite{Khvedelidze:2005rv}
\bibitem{Khvedelidze:2005rv}
  A.~Khvedelidze, A.~Kovner and D.~McMullan,
  %``Magnetic monopoles in 4D: A perturbative calculation,''
  arXiv:hep-th/0512142.
  %%CITATION = HEP-TH 0512142;%%
  %%Cited 0 times in SPIRES-HEP
  
%\cite{Smit:1993gy}
\bibitem{Smit:1993gy}
  J.~Smit and A.~J.~van der Sijs,
  %``Lattice computation of a magnetic monopole mass,''
  Int.\ J.\ Mod.\ Phys.\ C {\bf 5}, 347 (1994)
  [arXiv:hep-lat/9311045].
  %%CITATION = HEP-LAT 9311045;%%
  
  %\cite{Cea:2000zr}
\bibitem{Cea:2000zr}
  P.~Cea and L.~Cosmai,
  %``A gauge invariant study of the monopole condensation in non Abelian
  %lattice gauge theories,''
  Phys.\ Rev.\ D {\bf 62}, 094510 (2000)
  [arXiv:hep-lat/0006007].
  %%CITATION = HEP-LAT 0006007;%%
  
\bibitem{Ciria:1993yx}
  J.~C.~Ciria and A.~Tarancon,
  %``Renormalization group study of the soliton mass on the (lambda phi**4) in
  %(1+1)-dimensions lattice model,''
  Phys.\ Rev.\ D {\bf 49}, 1020 (1994)
  [arXiv:hep-lat/9309019].
  %%CITATION = HEP-LAT 9309019;%%
  
%\cite{Vettorazzo:2003fg}
\bibitem{Vettorazzo:2003fg}
  M.~Vettorazzo and P.~de Forcrand,
  %``Electromagnetic fluxes, monopoles, and the order of the 4d compact U(1)
  %phase transition,''
  Nucl.\ Phys.\ B {\bf 686} (2004) 85
  [arXiv:hep-lat/0311006].
  %%CITATION = HEP-LAT 0311006;%%
  
%\cite{Kajantie:1998zn}
\bibitem{Kajantie:1998zn}
  K.~Kajantie, M.~Laine, T.~Neuhaus, J.~Peisa, A.~Rajantie and K.~Rummukainen,
  %``Vortex tension as an order parameter in three-dimensional U(1) + Higgs
  %theory,''
  Nucl.\ Phys.\ B {\bf 546}, 351 (1999)
  [arXiv:hep-ph/9809334].
  %%CITATION = HEP-PH 9809334;%%
  
%\cite{Kajantie:2004vy}
\bibitem{Kajantie:2004vy}
  K.~Kajantie, M.~Laine, T.~Neuhaus, A.~Rajantie and K.~Rummukainen,
  %``Duality and scaling in 3-dimensional scalar electrodynamics,''
  Nucl.\ Phys.\ B {\bf 699}, 632 (2004)
  [arXiv:hep-lat/0402021].
  %%CITATION = HEP-LAT 0402021;%%
  
%\cite{Bogomolny:1975de}
\bibitem{Bogomolny:1975de}
  E.~B.~Bogomolny,
  %``Stability Of Classical Solutions,''
  Sov.\ J.\ Nucl.\ Phys.\  {\bf 24} (1976) 449
  [Yad.\ Fiz.\  {\bf 24} (1976) 861].
  %%CITATION = SJNCA,24,449;%%

%\cite{Prasad:1975kr}
\bibitem{Prasad:1975kr}
  M.~K.~Prasad and C.~M.~Sommerfield,
  %``An Exact Classical Solution For The 'T Hooft Monopole And The Julia-Zee
  %Dyon,''
  Phys.\ Rev.\ Lett.\  {\bf 35} (1975) 760.
  %%CITATION = PRLTA,35,760;%%
  
%\cite{Forgacs:2005vx}
\bibitem{Forgacs:2005vx}
  P.~Forgacs, N.~Obadia and S.~Reuillon,
  %``Numerical and asymptotic analysis of the 't Hooft-Polyakov magnetic
  %monopole,''
  Phys.\ Rev.\ D {\bf 71}, 035002 (2005)
  [Erratum-ibid.\ D {\bf 71}, 119902 (2005)]
  [arXiv:hep-th/0412057].
  %%CITATION = HEP-TH 0412057;%%
  
%\cite{Kirkman:1981ck}
\bibitem{Kirkman:1981ck}
  T.~W.~Kirkman and C.~K.~Zachos,
  %``Asymptotic Analysis Of The Monopole Structure,''
  Phys.\ Rev.\ D {\bf 24} (1981) 999.
  %%CITATION = PHRVA,D24,999;%%

%\cite{Gardner:1982fk}
\bibitem{Gardner:1982fk}
  C.~L.~Gardner,
  %``The 'T Hooft-Polyakov Monopole Near The Prasad-Sommerfield Limit,''
  Annals Phys.\  {\bf 146} (1983) 129.
  %%CITATION = APNYA,146,129;%%
  
%\cite{Kiselev:1990fh}
\bibitem{Kiselev:1990fh}
  V.~G.~Kiselev,
  %``A Monopole In The Coleman-Weinberg Model,''
  Phys.\ Lett.\ B {\bf 249}, 269 (1990).
  %%CITATION = PHLTA,B249,269;%%
    
%\cite{Luscher:1981zq}
\bibitem{Luscher:1981zq}
  M.~Luscher,
  %``Topology Of Lattice Gauge Fields,''
  Commun.\ Math.\ Phys.\  {\bf 85} (1982) 39.
  %%CITATION = CMPHA,85,39;%%
  
%\cite{Kronfeld:1990qu}
\bibitem{Kronfeld:1990qu}
  A.~S.~Kronfeld and U.~J.~Wiese,
  %``SU(N) Gauge Theories With C Periodic Boundary Conditions. 1. Topological
  %Structure,''
  Nucl.\ Phys.\ B {\bf 357} (1991) 521.
  %%CITATION = NUPHA,B357,521;%%
  
\bibitem{KleinertBook}
H.~Kleinert, {\it Gauge Fields in Condensed Matter} vol 1, (World Scientific, Singapore, 1989).

%\cite{Greensite:2004ke}
\bibitem{Greensite:2004ke}
  J.~Greensite, S.~Olejnik and D.~Zwanziger,
  %``Coulomb energy, remnant symmetry, and the phases of non-Abelian gauge
  %theories,''
  Phys.\ Rev.\ D {\bf 69}, 074506 (2004)
  [arXiv:hep-lat/0401003].
  %%CITATION = HEP-LAT 0401003;%%
  
%\cite{Baier:1988sc}
\bibitem{Baier:1988sc}
  R.~Baier, C.~B.~Lang and H.~J.~Reusch,
  %``The Renormalization Flow In The Adjoint SU(2) Lattice Higgs Model,''
  Nucl.\ Phys.\ B {\bf 305}, 396 (1988).
  %%CITATION = NUPHA,B305,396;%%  
  
%\cite{Peskin:1977kp}
\bibitem{Peskin:1977kp}
  M.~E.~Peskin,
  %``Mandelstam 'T Hooft Duality In Abelian Lattice Models,''
  Annals Phys.\  {\bf 113} (1978) 122.
  %%CITATION = APNYA,113,122;%%
  
    
\end{thebibliography}

\end{document}